\documentclass[aps,prl,twocolumn,superscriptaddress,amsmath,amssymb]{revtex4}

\usepackage{amssymb}
\usepackage{bm}
\usepackage{epsfig}
\usepackage[ansinew]{inputenc}
\usepackage{graphicx}
\usepackage{xcolor}

\begin{document}

\title{Absence of zero-field-cooled exchange bias effect in single crystalline La$_{2-x}$A$_{x}$CoMnO$_{6}$ (A = Ca, Sr) compounds}

\author{C. Macchiutti}
\affiliation{Centro Brasileiro de Pesquisas F\'{\i}sicas, Rio de Janeiro, RJ, 22290-180, Brazil}

\author{J. R. Jesus}
\affiliation{Centro Brasileiro de Pesquisas F\'{\i}sicas, Rio de Janeiro, RJ, 22290-180, Brazil}

\author{F. B. Carneiro}
\affiliation{Centro Brasileiro de Pesquisas F\'{\i}sicas, Rio de Janeiro, RJ, 22290-180, Brazil}

\author{L. Bufai\c{c}al}
\affiliation{Instituto de F\'{\i}sica, Universidade Federal de Goi\'{a}s, Goi\^{a}nia, GO, 74001-970, Brazil}

\author{M. Ciomaga Hatnean}
\affiliation{Department of Physics, University of Warwick, Coventry, CV4 7AL, UK}

\author{G. Balakrishnan}
\affiliation{Department of Physics, University of Warwick, Coventry, CV4 7AL, UK}

\author{E. M. Bittar}
\email{bittar@cbpf.br}
\affiliation{Centro Brasileiro de Pesquisas F\'{\i}sicas, Rio de Janeiro, RJ, 22290-180, Brazil}

\begin{abstract}
Magnetic properties of A$_{2}$BB'O$_{6}$ (A = rare or alkaline earth ions; B,B' = transition metal ions) double perovskites are of great interest due to their potential spintronic applications. Particularly fascinating is the zero field cooled exchange bias (ZEB) effect observed for the hole doped La$_{2-x}$A$_{x}$CoMnO$_{6}$ polycrystalline samples. In this work we synthesize La$_{2}$CoMnO$_{6}$, La$_{1.5}$Ca$_{0.5}$CoMnO$_{6}$, and La$_{1.5}$Sr$_{0.5}$CoMnO$_{6}$ single crystals by the floating zone method and study their magnetic behavior. The three materials are ferromagnetic. Surprisingly, we observe no zero or even conventional exchange bias effect for the Ca and Sr doped single crystals, in sharp contrast to polycrystalline samples. This absence indicates that the lack of grain boundaries and spin glass-like behavior, not observed in our samples, might be key ingredients for the spontaneous exchange bias phenomena seen in polycrystalline samples.
\end{abstract}

\maketitle

\section{Introduction}
Materials displaying general formula A$_{2}$BB'O$_{6}$ (A = rare or alkaline earth ions; B,B' = transition metal ions) and double perovskite structure present intriguing physical phenomena due to the interplay between structural, magnetic and electrical properties \cite{Vasala}. There is great interest in this family of compounds due to the possible applications in spintronic devices. Particularly interesting is the zero field cooled exchange bias (ZEB) effect recently observed for hole doped members of this family of compounds \cite{Murthy,Murthy2,LCCMO,CoMn_PRB,CoIr_PRB,CaCoMn_JMMM,Xie,Giri}. The conventional exchange bias (CEB) effect is characterized
as a shift in hysteresis loop measurements, observed after cooling the system in the presence of an external magnetic field \cite{CEB}. In materials in which the ZEB effect is seen, the shift in magnetization as a function of the applied field is detected spontaneously in zero field cooling \cite{Murthy,Murthy2,LCCMO,CoMn_PRB,CoIr_PRB,CaCoMn_JMMM,Xie,Giri}. In these double perovskites, the ZEB effect may result from a reentrant spin glass-like phase, in which magnetic relaxation of unusual glassy moments affects the magnetization loop shifts \cite{ZEBmodel,ZEBmodel2}.

The intrinsic magnetic inhomogeneity caused by different transition metal ions in double perovskites commonly leads to spin-glass-like behavior, making these materials good candidates to understand the mechanisms ruling the ZEB effect. The crystal structure of A$_{2}$BB'O$_{6}$ double perovskites is significantly influenced by the A, B, and B' cations and their ionic radii size mismatch. One can describe this mismatch by the Goldschmidt tolerance factor $t$, where $t=(r_A + r_O)/\sqrt{2}(<r_B> + r_O)$ and $r_A$, $<r_B>$, and $r_O$ are the ionic radii of A, average of B/B' and O, respectively \cite{Goldschmidt}. The ideal tolerance factor is given by $t=1$, which denotes the cubic rock-salt ordered, space group $Fm$-$3m$, structure. While $t<1$ indicates that the A cation ionic radii are smaller than ideal, and the structure compensates for its size mismatch by tilting the BO$_{6}$/B'O$_{6}$ octahedra. By decreasing slightly the tolerance factor the structure typically evolves to lower crystal symmetries as $Fm$-$3m$$\rightarrow$$I4/m$$\rightarrow$$R$-$3$$\rightarrow$$I2/m$$\rightarrow$$P2_1/n$ \cite{Vasala}. For ordered structures, a perfect sequence of B--O--B' chains are formed, whereas when the disorder is present, a random distribution of B and B' cations in which B--O--B, B--O--B' and B'--O--B' chains can occur. In disordered double perovskites, the ideal structure is the $Pm$-$3m$ space group, while the most common is $Pnma$ \cite{Vasala}.

The La$_{2-x}$A$_{x}$CoMnO$_{6}$ (A = Ca, Sr; $x$ = 0, 0.5) family of compounds have been studied over many years. The parent La$_{2}$CoMnO$_{6}$ (LCMO) material is known to show ferromagnetic behavior due to superexchange Co$^{2+}$--O--Mn$^{4+}$ coupling, according to Goodenough-Kanamori-Anderson (GKA) rules \cite{GKA,Blasse}, with potential multiferroic properties \cite{Singh,Chen}. The Curie temperature ($T_C$) for ordered samples can reach up to 226 K \cite{Dass} with a saturation spin moment $M_{\rm sat}=6$ $\mu$B/f.u., while the disordered ones, \textit{e.g.} with oxygen deficiency and/or antisite defects, present lower $T_C$ and $M_{\rm sat}$ \cite{Murthy,Dass,Bull}. If changes in the B-cations valence also occur, a Co$^{3+}$--O--Mn$^{3+}$ ferromagnetic interaction may happen, and a distinct ferromagnetic transition is present \cite{CoMn_PRB,GKA,Chen,Dass}.  Also, the disorder can generate antiferromagnetic clusters, giving rise to an antiferromagnetic transition temperature $T_N$ because of exchange Co$^{3+}$--O--Mn$^{4+}$ coupling and/or antiphase boundaries of Co$^{2+}$--Co$^{2+}$ or Mn$^{4+}$--Mn$^{4+}$ pairs \cite{GKA,Bai}. Regarding its crystal structure, LCMO is mostly described as a monoclinic $P2_1/n$ space group for an ordered population of the B-sites, and orthorhombic $Pnma$ when the disorder is present \cite{CoMn_PRB,Dass}.

For La$_{1.5}$Ca$_{0.5}$CoMnO$_{6}$ (LCCMO) and La$_{1.5}$Sr$_{0.5}$CoMnO$_{6}$ (LSCMO) materials, it has been reported in polycrystalline samples that by substituting the La$^{3+}$ cation by 25\% of an alkaline earth ion with 2+ valence, gives rise to a spin glass phase which promotes the unusual ZEB effect \cite{Murthy,Murthy2,CoMn_PRB,CaCoMn_JMMM,ZEBmodel,ZEBmodel2}.  It was also shown that Ba doping can tune the ZEB effect in LSCMO \cite{APL}. In polycrystalline LCCMO, a phase segregation with mixture of two $Pnma$ space groups in 94\% and 6\% concentrations was seen, while LSCMO showed 91\% of rhombohedral $R3$-$c$ and 9\% of $Pnma$ phases \cite{CoMn_PRB}. Magnetic properties of both samples showed two ferromagnetic ($T_C$) and one antiferromagnetic ($T_N$) transition temperatures, $T_{C1}\sim158$ K, $T_{C2}\sim141$ K and $T_{N}\sim62$ K, for LCCMO; and $T_{C1}\sim180$ K, $T_{C2}\sim157$ K and $T_{N}\sim74$ K, for LSCMO . Both polycrystalline compounds exhibit the onset of cluster glass behavior at lower temperatures due to the presence of competing magnetic phases \cite{CoMn_PRB}.

Since the CEB is an interface effect \cite{CEB}, it is usually observed in polycrystalline core-shell systems or heterostructured thin films presenting two or more magnetic phases. Similarly, for the ZEB effect, most studies in the literature have focused on polycrystalline materials. In single crystals, however, the reports of the ZEB phenomenon are scarce. There is an intersection between different single crystalline perovskites compounds, such as La$_{0.82}$Sr$_{0.18}$CoO$_{3}$ \cite{Huang2008}, Y$_{0.95}$Eu$_{0.05}$MnO$_{3}$  \cite{ Xiao2016}, and SmFeO$_{3}$ \cite{Fita2018,Wang2018,Ding2019}, that show an essential role of a spin glass phase for the observance of the ZEB effect in these materials. Thus, additional research in similar systems is desired to understand the phenomenon thoroughly.

Here we present our study on single crystalline double perovskite LCMO, LCCMO, and LSCMO compounds, where we investigate their structural and magnetic properties, comparing their characteristics with the polycrystals. All samples are ferromagnetic below $T_C$, with a clear indication of two ferromagnetic transitions for LCMO and LCCMO. Changes in $T_C$ between these oxides are accounted for structural modifications of the B$-$O$-$B bound angle, B$-$O bond length, and crystallinity. We observe no glassy-like phase for the three materials and the absence of the ZEB effect for LCCMO and LSCMO. Therefore, the nonexistence of grain boundaries and spin glass-like behavior indicate that these are critical ingredients for the spontaneous exchange bias phenomena seen in polycrystalline samples.

\section{Experimental details}

Polycrystalline powders of LCMO, LCCMO, and LSCMO materials were synthesized via solid-state reaction, details of the process have been reported elsewhere \cite{CoMn_PRB}.  The crystal growths were carried out by the floating zone technique using two different 2-mirror image furnaces (NEC SC1MDH-11020 Canon Machinery Inc. and Quantum Design IR Image Furnace), in air. The feed rods needed for the crystal growths were hydrostatically pressed up to 50 MPa into the form of cylindrical rods and sintered in air at 1300$^\circ$C for 12 h (LCMO), 1300$^\circ$C for 24 h (LCCMO), and 1400$^\circ$C for 24 h (LSCMO). Crystals were grown at a rate of 3 to 5 mm/h with the feed and seed rods, each rotating at 20 rpm in opposite directions. The structural properties of the samples were investigated by conventional x-ray powder diffraction (XRPD) measurements using a PANalytical Empyrean diffractometer with Cu-$K_\alpha$ radiation ($\lambda=1.5406$ \AA). All data were collected in Bragg-Brentano geometry in the continuous mode with 2$\theta$ range from 20$^\circ$ to 80$^\circ$, step size of 0.013$^\circ$, and a scanning speed of 0.5$^\circ$/min. Rietveld refinements were performed using the GSAS+EXPGUI suite \cite{GSAS+EXPGUI}. X-ray Laue back-scattering patterns were recorded using a commercial Photonic Science Laue System, for cutting and aligning the samples along different directions for magnetic measurements. Roughly, 5 mm size single crystals were extracted from the crystal boules. Magnetic and thermal properties characterization was carried out using a Quantum Design PPMS equipment.

\section{Results}

\subsection{Structural characterization}

\begin{figure}
\begin{center}
\includegraphics[width=0.4\textwidth]{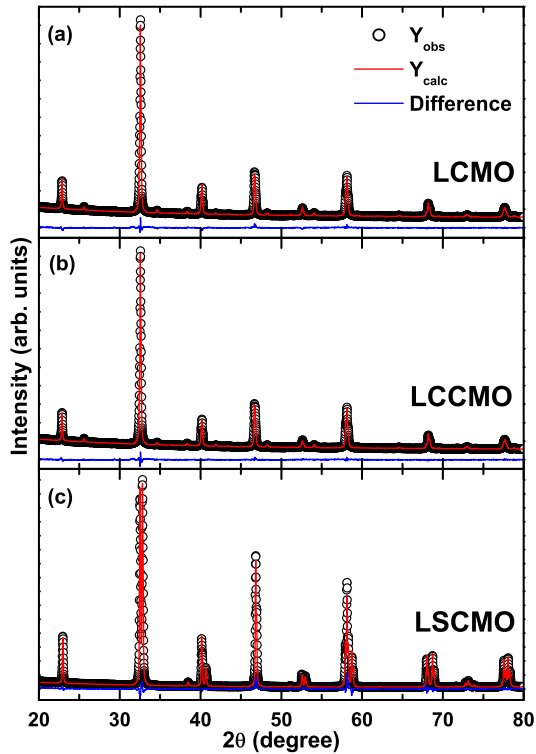}
\end{center}
\caption{Conventional Cu-$K_\alpha$ x-ray powder diffraction patterns for (a) LCMO, (b) LCCMO and (c) LSCMO. Open black circles are experimental data; red curves calculated Rietveld refinement; blue curves the difference between the observed and calculated patterns.}
\label{Fig1}
\end{figure}

Figure \ref{Fig1} shows the XRPD patterns collected on LCMO, LCCMO, and LSCMO samples. These powdered samples were obtained by powdering a small section of the as-grown boule and measured at room temperature. For the three samples all the Bragg peaks could be indexed using the double perovskite structure model and no impurity peaks could be observed. The crystal structure was indexed with orthorhombic symmetry, space group $Pnma$, for the LCMO and LCCMO; while LSCMO is rhombohedral, space group $R$-$3c$. Lattice parameters (see Table \ref{T_XRD}) are in good agreement with the data reported for polycrystalline samples on literature \cite{CoMn_PRB}.

\begin{table}
\caption{Main results obtained from the Rietveld refinements of the XRPD data.}
\label{T_XRD}
\centering
\begin{tabular}{ccccc}
\hline \hline
 & LCMO & LCCMO & LSCMO \\
\hline
Space Group & $Pnma$ & $Pnma$ & $R$-$3c$  \\
$a$ (\AA) & 5.4699(3) & 5.4908(1) & 5.5130(1)  \\
$b$ (\AA) & 7.7182(3) & 7.7668(1) & 5.5130(1) \\
$c$ (\AA) & 5.4607(2) & 5.5121(1) & 13.3183(1) \\
$<$B$-$O$>$ (\AA) & 1.9520(12) & 1.9806(1) & 1.9581(3) \\
$<$B$-$O$-$B$>$ ($^{\circ}$) & 162.6(11) & 159.2(4) & 164.5(1) \\
$R_p$ (\%) & 3.8 & 1.2 & 3.5 \\
$R_{wp}$ (\%) & 5.7 & 1.8 & 4.7 \\
\hline \hline
\end{tabular}
\end{table}

\begin{figure}
\begin{center}
\includegraphics[width=0.5\textwidth]{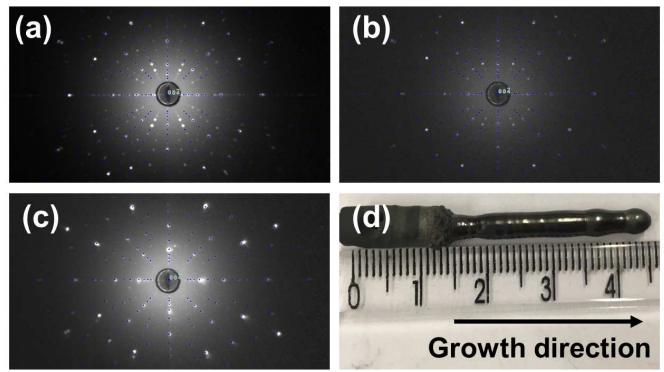}
\end{center}
\caption{X-ray Laue back reflection photograph, showing the [001] orientation of an aligned sample piece used for physical properties measurements, for (a) LCMO, (b) LCCMO, and (c) LSCMO. (d) LCCMO as-grown single crystal boule grown by the floating zone method.}
\label{Fig2}
\end{figure}

Figure \ref{Fig2} shows the x-ray Laue back-scattering patterns of the three LCMO, LCCMO, and LSCMO samples, after the as-grown single crystal boules [Fig. \ref{Fig2}(d)] were oriented and cut along the [001] crystalline direction. Fig. \ref{Fig2}(b) presents better-defined diffraction spots resulting from a more homogeneous sample. It indicates that the proximity between La$^{3+}$ and Ca$^{2+}$ ionic radii leads to a better accommodation of these ions in the LCCMO structure than LSCMO resulting from the incorporation of large Sr$^{2+}$ ions. The LCCMO has less residual stress and better crystal quality manifested in the goodness of fit parameters obtained from the Rietveld refinements (see Table \ref{T_XRD}).

\subsection{Magnetic properties}

\begin{figure}
\begin{center}
\includegraphics[width=0.5\textwidth]{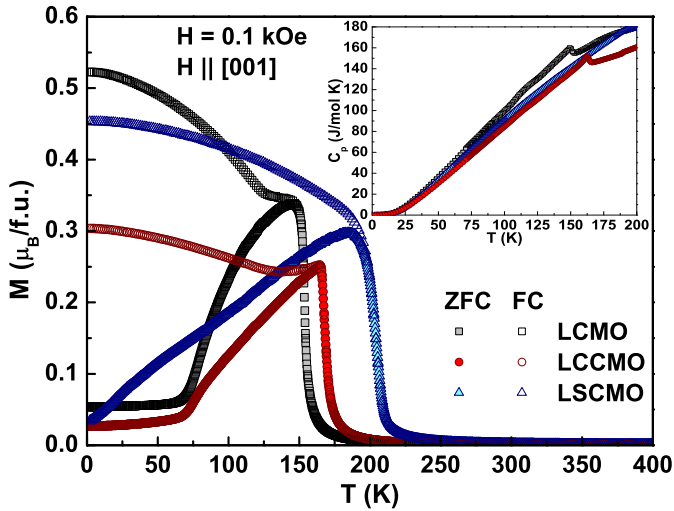}
\end{center}
\caption{LCMO, LCCMO and LSCMO as-grown single crystals temperature dependence of the zero field cooled (ZFC) and field cooled (FC) magnetization curves for $H$ $||$ [001], measured at $H=0.1$ kOe. Inset: Zero magnetic field heat capacity as a function of temperature.}
\label{Fig3}
\end{figure}

The zero field cooled (ZFC) and field cooled (FC) temperature dependent magnetization of the LCMO, LCCMO, and LSCMO materials are shown in Fig. \ref{Fig3}, under an applied magnetic field ($H$) of 0.1 kOe along the [001] direction. No appreciable qualitative changes were observed for $H \perp$ [001] for all samples (not shown). The magnetic irreversibility of the ZFC and FC curves observed at the paramagnetic to ferromagnetic transition for LCMO, LCCMO, and LSCMO is usually ascribed to a glassy-like magnetic behavior. Especially for LCMO and LCCMO, the ZFC magnetization only starts to grow at around $T=75$ K.

\begin{figure}
\begin{center}
\includegraphics[width=0.5\textwidth]{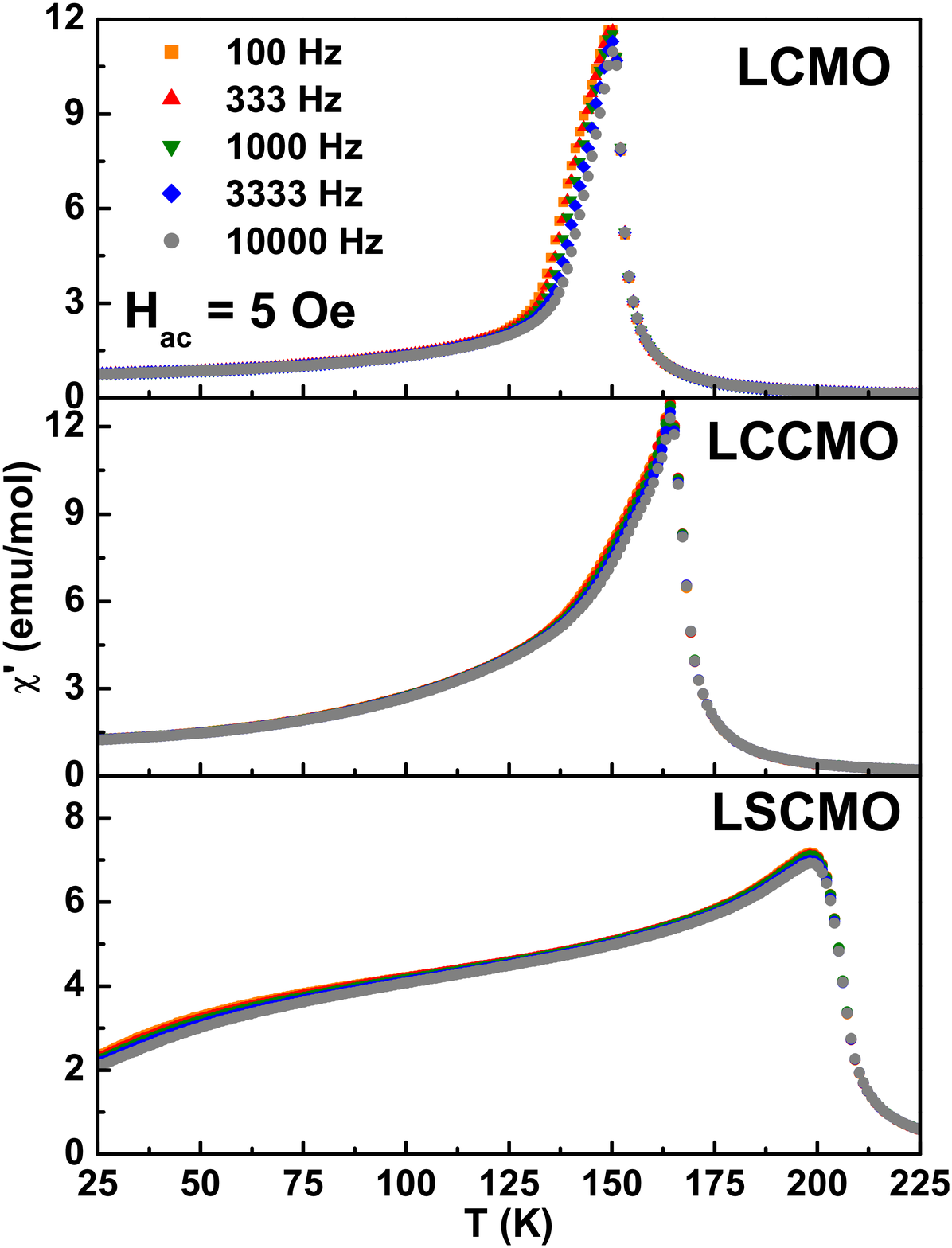}
\end{center}
\caption{LCMO, LCCMO, and LSCMO single crystals in-phase component of ZFC \textit{ac} susceptibility ($\chi'$) as a function of temperature, for different frequencies in a probing \textit{ac} magnetic field of 5 Oe.}
\label{Fig4}
\end{figure}

In order to investigate this hypothesis, were performed measurements of \textit{ac} susceptibility in-phase component ($\chi'$) as a function of temperature, as shown in Fig. \ref{Fig4}. Although some changes in the intensities of the \textit{ac} curves could be noticed in the low temperature side of the peaks carried at different frequencies, no systematic variations in the temperature of the peaks were observed, indicating no canonical glassy phase nor relevant dynamic effects \cite{SG}. Conversely, this indicates that such frequency dependence of the \textit{ac} intensity is most likely related to the onset of a second ferromagnetic transition extended over a broad temperature range, as will be discussed in the next section. The magnetic irreversibility observed can be attributed to the domain wall motion de-pinning process, as previously seen in other double perovskites single crystals \cite{SCLCMO,YCMO}. Table \ref{T2} summarises the magnetic ordering transition temperatures for all studied materials. We could not obtain the effective moment nor the Curie-Weiss temperature due to limitations of the maximum temperature measured ($T=400$ K), preventing a good linear fit in the paramagnetic region. Nevertheless, this is an indication of short-range magnetic correlations even at temperatures well above the ordering temperatures.

\begin{table}
\caption{Main results obtained from the temperature dependence of the magnetization and ZFC $M(H)$ curves.}
\label{T2}
\centering
\begin{tabular}{cccc}
\hline \hline
Sample & LCMO & LCCMO & LSCMO \\
\hline
$T_{C}$ (K) & 150(1) & 165(2) & 205(2)\\
$T_{C}'$ (K) & 116(2) & 120(2) &  \\
$H_C$ (Oe) & 1396(116) & 626(112) & 1380(88) \\
$M_S$ ($\mu_B$/f.u.) & 6.3(1) & 6.8(1) & 6.5(1) \\
$M_R$ ($\mu_B$/f.u.) & 2.4(1) & 0.7(1) & 2.7(1) \\
\hline \hline
\end{tabular}
\end{table}

\begin{figure}
\begin{center}
\includegraphics[width=0.5\textwidth]{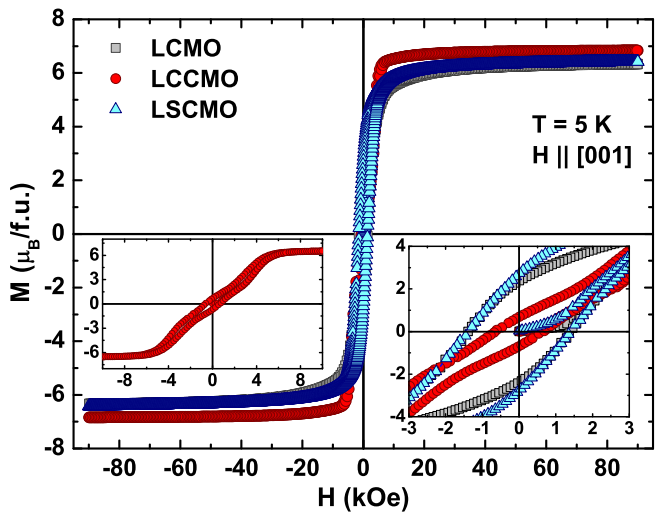}
\end{center}
\caption{LCMO, LCCMO and LSCMO ZFC as-grown single crystals $M(H)$ curves for $H$ $||$ [001], measured at $T=5$ K and $H_{max}=90$ kOe. Right inset
shows magnified view of the curves close to the $M=0$ region. Left inset LCCMO $M(H)$ curve.}
\label{Fig5}
\end{figure}

Figure \ref{Fig5} shows the magnetization hysteresis loop at $T=5$ K and up to $H=90$ kOe for LCMO, LCCMO, and LSCMO. These curves were obtained after zero field cooling the samples. A sharp ferromagnetic-like hysteresis is seen for all three samples, and the magnetization saturates at $M_S\sim6$ $\mu_B$/f.u for $H>40$ kOe. We see no hysteresis loop shifts, \textit{i.e.}, no ZEB effect, for LCCMO and LSCMO single crystals (see right side inset Fig. \ref{Fig5}), in contrast to what was observed in polycrystals \cite{Murthy,Murthy2,Giri,LCCMO,CaCoMn_JMMM,CoMn_PRB}. Also, we do not observe the conventional EB effect, after FC at $H=9$ T, for all samples (not shown). In addition, no appreciable changes for the magnetization hysteresis loop either at ZFC or FC at $H=9$ T for the other orthogonal crystalline directions were seen (not shown). A metamagnetic transition can be noticed for LCCMO (see left side inset Fig. \ref{Fig5}).

\section{Discussion}

\subsection{LCMO}

Regarding the magnetic properties, for LCMO, we observe two ferromagnetic transitions, a majority ferromagnetic phase at $T_{C}\sim150(1)$ K and a minority ferromagnetic phase at $T_{C}'\sim116(2)$ K, as seen from the FC curve in Fig. \ref{Fig3}. $T_{C}$ is similar as reported before for single crystals \cite{SCLCMO,SCLCMO2}. These magnetic phase transitions are supported by zero magnetic field heat capacity measurement, showing one broad peak at approximately 116 K and another sharper at around 150 K (see inset of Fig. \ref{Fig3}). In polycrystals it was seen by x-ray absorption spectroscopy measurements the presence of mixed valence Co (Co$^{2+}$/Co$^{3+}$) and Mn (Mn$^{3+}$/Mn$^{4+}$) \cite{CoMn_PRB,XASLCMO}. Thus, by the GKA rules \cite{GKA}, one can ascribe these ferromagnetic orderings as result of Co$^{2+}$--O--Mn$^{4+}$ ($T_{C}$) and Co$^{3+}$--O--Mn$^{3+}$ ($T_{C}'$) superexchange interactions, with $T_{C}>T_{C}'$ \cite{Dass,CoMn_PRB,XASLCMO}. The presence of Co$^{3+}$ and Mn$^{3+}$ was already reported for LCMO solution grown single crystals \cite{SCLCMO2}. The \textit{ac} susceptibility (Fig. \ref{Fig4}) shows a frequency-independent peak, within the experimental resolution, approximately at 150 K. However, no peak is observed at around 116 K. This indicates that at $T_{C}$ a long-range magnetic order sets in, while at $T_{C}'$ the Co$^{3+}$--O--Mn$^{3+}$ exchange interaction has more a short-range character. The small frequency dependence for $125<T<150$ K indicates slow spin dynamics, which is not characteristic of a canonical glassy behavior and may be due to domain-wall motion.

Figure \ref{Fig5} shows a sharp ferromagnetic magnetization hysteresis loop (up to $H=90$ kOe) which saturates at $M_S=6.3(1)$ $\mu_B$/f.u. and has remanent magnetization $M_R=2.4(1)$ $\mu_B$/f.u. (see Table \ref{T2}). For the fully ordered LCMO one would expect only the Co$^{2+}$--O--Mn$^{4+}$ exchange interaction and $M_S=6$ $\mu_B$/f.u, consistent with the sum of Co$^{2+}$ ($3d^7$, $S=3/2$) and Mn$^{4+}$ ($3d^3$, $S=3/2$) states per formula unit \cite{Dass}. Since the temperature dependence of magnetization of our single crystalline sample already gave us an indication of mixed valence, we must take into account Co$^{3+}$ and Mn$^{3+}$ spins to the saturated experimental value. Considering, high spin Co$^{3+}$ ($3d^6$, $S=2$), and Mn$^{3+}$ ($3d^4$, $S=2$), one can estimate the amount of Co$^{2+}$--O--Mn$^{4+}$ and Co$^{3+}$--O--Mn$^{3+}$ couplings, by equating the expression $M_S=6.3=[(1-x)($Co$^{2+}+$Mn$^{4+})+x($Co$^{3+}+$Mn$^{3+})]$, thus deriving that our LCMO sample has 15\% of Co$^{3+}$--O--Mn$^{3+}$ and 85\% of Co$^{2+}$--O--Mn$^{4+}$ interactions. In this rough approximation, we are neglecting the cationic disorder-induced contributions of antiferromagnetic couplings, such as Mn$^{4+}$--Mn$^{4+}$, Co$^{2+}$--Mn$^{3+}$, Co$^{2+}$--Co$^{2+}$. And also ferromagnetic couplings, for example, Mn$^{3+}$--Mn$^{4+}$. This consideration is a good approximation once $H=90$ kOe may be sufficient to flip the great majority of the antiferromagnetic spins pointing toward the opposite direction to the external magnetic field.

\subsection{LCCMO}

By hole doping Ca$^{2+}$ and Sr$^{2+}$ into the La$^{3+}$ site in the LCMO structure, one destabilizes the electron count, and in order to maintain charge neutrality, it is more likely that either some Mn$^{3+}$ becomes Mn$^{4+}$ or Co$^{2+}$ transforms to Co$^{3+}$. Previous x-ray absorption spectroscopy measurements showed indications of a preference for the increase of the amount of Co$^{3+}$ by alkaline-earth cations substitution in polycrystalline LCMO. At the same time, Mn is maintaining its mean valence value \cite{CoMn_PRB}.

For LCCMO we observe a clear ferromagnetic transition at $T_{C}\sim165(2)$ K (Figs. \ref{Fig3} and \ref{Fig4}), slightly greater than for LCMO. It is also noticed a broader second transition at $T_{C}'\sim120$ K. The $M_S=6.8(1)$ $\mu_B$/f.u. seen in Fig. \ref{Fig5} is also greater for LCCMO than for LCMO. Both results are due to the doping-induced electronic changes. Assuming that the Mn mean valence does not change with Ca doping, there should be an increase of 50\% of Co$^{3+}$ in respect to LCMO. One can calculate the expected saturation magnetization by the expression $M_S=[0.85$Mn$^{4+}+0.15$Mn$^{3+}+(0.85-0.50)$Co$^{2+}+(0.15+0.50)$Co$^{3+}]$, which gives $M_S=6.8$ $\mu_B$/f.u., exactly the obtained experimental value. The increased disorder caused by changes in the valence of the transition metal ions prevents the percolation of the Co$^{3+}$--O--Mn$^{3+}$ coupling. This transition might be smeared over a wide temperature range (see inset Fig. \ref{Fig3}) as seen for the frequency dependence of $\chi'$ at $125<T<150$ K (Fig. \ref{Fig4}).

Left inset of Fig. \ref{Fig5} highlights a step-like character in the $M(H)$ curve of LCCMO at $T\sim5$ K. Such behavior was reported for similar compounds, in which different mechanisms were said to explain such effect. For Tb$_2$CoMnO$_6$ and Eu$_2$CoMnO$_6$ this step-like behavior was initially explained in terms of a field-induced metamagnetic transition from ferrimagnetism to ferromagnetism in the Co$^{2+}$--O--Mn$^{4+}$ coupling \cite{EuCoMn}. Nevertheless, later on a more recent detailed investigation employing neutron powder diffraction ruled out such possibility \cite{TbCoMn2}. In this study, the metamagnetic transition was explained in terms of the presence of a spin-glass-like phase and of the magnetic ordering of rare-earth ions in a direction distinct to those of Co, and Mn \cite{TbCoMn2,TbCoMn}. In our case, in contrast to these hypotheses, the A-site ions are diamagnetic, and no sign of a spin-glass-like phase is observed. Thus, the characteristic feature observed in the $M(H)$ curve of LCCMO may be justified in terms of the two distinct ferromagnetic couplings contributions for the total magnetization.

\subsection{LSCMO}

The LSCMO sample present $T_C\sim205(2)$ K (Table \ref{T2}), and a second ferromagnetic/antiferromagnetic transition at lower temperatures is not clear, in contrast to polycrystals \cite{Murthy2,CoMn_PRB}. As in LCCMO, one would expect an increase of the Co valence, though, for LSCMO, the saturated magnetization is slightly smaller ($M_S=6.5(1)$ $\mu_B$/f.u.), indicating that not only Co but also Mn valence changes. If we assume that in LSCMO, all 15\% of Mn$^{3+}$ becomes Mn$^{4+}$, then only 35\% of Co$^{2+}$ shifts into Co$^{3+}$. Thus, the equation $M_S=[(0.85+0.15)$Mn$^{4+}+(0.15-0.15)$Mn$^{3+}+(0.85-0.35)$Co$^{2+}+(0.15+0.35)$Co$^{3+}]=6.5$ $\mu_B$/f.u., has the same value that the experimental magnetization saturates (Fig. \ref{Fig5}). However, the weak frequency dependence in the in-phase component of the \textit{ac} susceptibility at low temperatures indicates an almost negligible amount of Mn$^{3+}$ may be present. In such a case, the diluted Co$^{3+}$--Mn$^{3+}$ interaction would be then pushed to lower temperatures in respect to LCMO and LCCMO (Fig. \ref{Fig4}).

The ferromagnetic transition temperature $T_{C}$ increases for doped LCCMO and LSCMO regarding the undoped LCMO. Possible explanations to this, which might be counterintuitive, once doping causes disorder and also significantly promotes some Co$^{2+}$ to Co$^{3+}$, may come from the crystal structure properties of these A$_{2}$BB'O$_{6}$ materials. The effect of the transition metal exchange interaction (B$-$O$-$B) orbital overlapping geometry is more critical to strengthen $T_{C}$ than disorder is to weaken it \cite{Bai}. Therefore, one suitable structural parameter for this analysis is the B$-$O$-$B bound angle and the B$-$O bond length. The closer the bond angle is to 180$^{\circ}$ and the shorter is the bond length, the higher is $T_{C}$.

\subsection{Conclusions}

Considering the ionic radii of the following elements as being equal to La$^{3+}$ (1.36 \AA), Ca$^{2+}$ (1.34 \AA), Sr$^{2+}$ (1.44 \AA) [XII coordination]; Co$^{2+}$ (0.745 \AA), Co$^{3+}$ (0.61 \AA), Mn$^{3+}$ (0.645 \AA), Mn$^{4+}$ (0.53 \AA) [VI coordination]; and O$^{2-}$ (1.35 \AA) [II coordination] \cite{iRadii}, one can calculate the tolerance factor for each material as defined in the introduction. The stable perovskite structure is guaranteed by $0.89<t<1.02$ and as $t$ decreases, the crystalline lattice becomes rhombohedral ($0.96<t<1$) and orthorhombic ($t <0.96$) \cite{Goldschmidt2}. As in previous polycrystalline samples \cite{CoMn_PRB}, the parent LCMO compound has orthorhombic symmetry. The calculated tolerance factor for this material is $t\sim0.96$, in agreement with the observed $Pnma$ space group. For LCCMO, we see a decrease of the bond angle and an increase of the bond length (see Table \ref{T_XRD}). The calculated $t\sim0.98$ indicates a preference for a rhombohedral atomic arrangement, though experimentally, we find the orthorhombic structure. However, the greater structural distortion due to doping gives rise to the tilting of the oxygen octahedra and accommodates the ions better in the crystal lattice, explaining the higher crystal homogeneity of the LCCMO sample when compared to LCMO and LSCMO [see Fig. 1(a) and Fig. 2(b) for sharper diffraction peaks and spots, respectively]. This higher degree of crystallinity could describe the small increase in $T_{C}$, despite reduced bond angle and elongated bond length regarding LCMO. The calculated $t\sim0.99$ for LSCMO produces a more symmetric rhombohedral crystal structure, and the bond angle is closer to 180$^{\circ}$ giving rise to the highest $T_{C}$.

Despite many works on the ZEB effect in polycrystalline Ca$^{2+}$ and Sr$^{2+}$ doped LCMO, our single crystals showed no signatures of such effect (right inset Fig. \ref{Fig5}). Neither did we observe the conventional magnetization as a function of applied field shift by cooling the materials under an external magnetic field ($H=9$ T cooling field; not shown). One key difference between poly and single crystals is the lack of grain boundaries in the latter, in which the number of interfaces is drastically reduced. It is well established that the exchange bias effect is an interface phenomenon between different magnetic phases, usually a ferromagnetic/antiferromagnetic uncompensated moment interface \cite{CEB}. Thus, in a single magnetic phase single crystal, it is expected that no conventional exchange bias effect is to be observed. The presence of a reentrant magnetic glassy-like phase would, however, change this scenario. The slow dynamics of the spin glass-like moments relaxation under a magnetic field would create an artificial interface of pinned moments, giving rise to the asymmetry in the hysteresis loops \cite{CoMn_PRB,Wang,Maity,Nayak}. In our LCCMO and LSCMO samples, we see no such glassy-like phase. Therefore, the absence of the ZEB effect in single crystals attests to the need for a glassy magnetic phase for its insurgence in these types of materials \cite{ZEBmodel,ZEBmodel2}. In addition, we see some probable correlation of the ZEB effect magnitude with the grain boundary size in polycrystals, and further studies should verify this.

\section{Summary}

Here we present our work on the magnetic properties of single crystalline La$_{2-x}$A$_{x}$CoMnO$_{6}$ (A = Ca, Sr; $x$ = 0, 0.5) materials. The crystalline structure of the single crystals is the same as found for polycrystalline samples \cite{CoMn_PRB}, orthorhombic for LCMO and LCCMO, and rhombohedral for LSCMO. LCMO and LCCMO present two magnetic transitions ascribed to the Co$^{2+}$--O--Mn$^{4+}$ and Co$^{3+}$--O--Mn$^{3+}$ ferromagnetic couplings. For LSCMO, however, only one $T_C$ is clear, most likely due to Co$^{2+}$--O--Mn$^{4+}$ long-range interaction. The increased $T_C$ of LCCMO (165 K) and LSCMO (205 K) in relation to LCMO (150 K) is primarily because of structural changes in crystallinity and bond angles and lengths. No zero or even conventional exchange bias effect is seen for the Ca and Sr doped single crystals, in sharp contrast to polycrystalline samples \cite{Murthy,Murthy2,LCCMO,CoMn_PRB}. The absence of the ZEB effect might be due to the lack of a spin glass-like phase and/or grain boundary interfaces in these single crystals. Further studies on the spin dynamics may elucidate the nonexistence of the ZEB effect in single crystalline samples and give insights into the critical properties necessary for its appearance.

\begin{acknowledgements}
This work was supported by the Brazilian funding agencies: Funda\c{c}\~{a}o Carlos Chagas Filho de Amparo \`{a} Pesquisa do Estado do Rio de Janeiro (FAPERJ) [Nos. E-26/202.798/2019], Funda\c{c}\~{a}o de Amparo \`{a}  Pesquisa do Estado de Goi\'{a}s (FAPEG) and Conselho Nacional de Desenvlovimento Cient\'{\i}fico e Tecnol\'{o}gico (CNPq) [No. 400633/2016-7]. The work at the University of Warwick was funded by EPSRC, UK, through Grant EP/T005963/1. 
\end{acknowledgements}

\end{document}